\documentclass[preprint,showpacs,preprintnumbers,amsmath,amssymb]{revtex4}

\pdfoutput=1

\usepackage{graphicx}
\usepackage{dcolumn}
\usepackage{bm}
\usepackage{amssymb}

\begin{document}
\title{
Validity of the time-dependent variational approximation to
the Gaussian wavepacket method applied to double-well systems
}

\author{Hideo Hasegawa}
\altaffiliation{hideohasegawa@goo.jp}
\affiliation{Department of Physics, Tokyo Gakugei University,  
Koganei, Tokyo 184-8501, Japan}%

\date{\today}
\begin{abstract}
We have examined the validity of the time-dependent variational approximation (TDVA)
to the Gaussian wavepacket method (GWM) for quantum double-well (DW) systems, 
by using the quasi-exact spectral method (SM).
Comparisons between results of wavefunctions, averages of position and momentum, 
the auto-correlation function, and an uncertainty product calculated by SM and TDVA
have been made. It has been shown that a given initial Gaussian wavepacket in SM 
is quickly deformed at $t > 0$ where a wavepacket cannot be expressed 
by a {\it single} Gaussian, and that assumptions on averages of
higher-order fluctuations in TDVA are not justified.
These results cast some doubt on an application of TDVA to DW systems.
Gaussian wavepacket dynamics in anharmonic potential systems is studied also.

\vspace{0.5cm}

\end{abstract}

        

\maketitle
\newpage
\section{Introduction}
Dynamical properties of nonrelativistic quantum systems may be described
by the Schr\"{o}dinger equation \cite{Tannor07}, in which the 
time-dependent wavefunction $\Psi(x,t)$ 
for the one-dimensional system with the potential $U(x)$ is described by 
\begin{eqnarray}
i \hbar \:\frac{\partial \Psi(x,t)}{\partial t} &=& H \Psi(x,t) 
= \left[ - \frac{\hbar^2}{2m}\frac{\partial^2}{\partial x^2}+ U(x) \right]\Psi(x,t).
\label{eq:A1}
\end{eqnarray}
It is generally difficult to obtain exact solutions of the Schr\"{o}dinger equation 
which are available only for limited cases like a harmonic oscillator (HO) system.
For general quantum systems, various approaches such as perturbation and spectral methods 
have been developed to obtain approximate solutions \cite{Tannor07}.
From Eq. (\ref{eq:A1}),
we may derive equations of motion for $\langle x \rangle$ and $\langle p \rangle$ 
expressed by
\begin{eqnarray}
\frac{d \langle x \rangle}{dt}  &=& \frac{\langle p \rangle}{m}, \;\;\;\;
\frac{d \langle p \rangle}{dt} = - \left< \frac{\partial U(x)}{\partial x} \right>,
\label{eq:A2}
\end{eqnarray}
where the bracket $\langle \cdot \rangle$ denotes the expectation value.
Although equations of motion given by Eq. (\ref{eq:A2}) are closed 
within $\langle x \rangle$ and $\langle p \rangle $ for a HO system, they generally yield 
equations of motion including higher-order fluctuations such as $\langle \delta x^2 \rangle$,
$\langle \delta p^2 \rangle$ and 
$\langle \delta x \delta p+ \delta p \delta x \rangle$ where
$\delta x=x-\langle x \rangle$ and $\delta p=p-\langle p \rangle$.
It is necessary to develop an approximate method to close or truncate
a hierarchical chain of equations of motion.

The Gaussian wavepacket method (GWM) is one of such methods
whose main aim is a semi-classical description of quantum systems
(for a recent review on GWM, see Ref. \cite{Robinett04}).
If the wavefuction is Gaussian at $t=0$ in a HO system, it remains
at all $t > 0$.
Heller \cite{Heller75} proposed that even for more realistic potentials,
we may adopt a (thawed) Gaussian wavepacket given by
\begin{eqnarray}
\Psi_H(x, t) &=& \exp\left[ \frac{i}{\hbar}
[A(x-\langle x \rangle)^2+ \langle p \rangle (x-\langle x \rangle)+ \gamma] \right],
\label{eq:A3}
\end{eqnarray}
where 
$A$ and $\gamma$ are time-dependent complex parameters.
Heller \cite{Heller75} derived equations of motion for $\langle x \rangle$, 
$\langle p \rangle$, $A$ and $\gamma$,
employing an assumption that
the potential expanded in the Taylor series at $x=\langle x \rangle$ may be truncated by
\begin{eqnarray}
U(x) &\cong& U^{(0)}(\langle x \rangle)+U^{(1)}(\langle x \rangle)(x-\langle x \rangle)
+ \frac{1}{2}U^{(2)}(\langle x \rangle)(x-\langle x \rangle)^2,
\label{eq:A4}
\end{eqnarray}
where $U^{(k)}(x)$ signifies the $k$th derivative of $U(x)$.
The concept of the Gaussian wavepacket has been adopted in many fields \cite{Robinett04}.
Dynamics is well described by GWM for a HO system where motions of fluctuations 
are separated from those of $\langle x \rangle$ and $\langle p \rangle$,
leading to the uncertainty relation: 
$\langle \delta x^2\rangle \langle \delta p^2\rangle \geq \hbar^2/4$.
Various types of variants of GWM such as the frozen \cite{Heller81} 
and generalized Gaussian wavepacket methods \cite{Huber87} have been proposed \cite{Robinett04}. 
Among them, we pay our attention into the time-dependent variational approximation (TDVA) 
which employs the normalized squeezed coherent-state Gaussian wavepacket given by 
\cite{Cooper86,Tsue91,Pattanayak94,Pattanayak94b,Sundaram95}
\begin{eqnarray}
\Psi_G(x,t)=\frac{1}{(2 \pi \mu)^{1/4}} 
\:\exp\left[-\frac{(1-i \alpha)}{4 \mu}(x-\langle x \rangle)^2
+ i \:\frac{\langle p \rangle (x- \langle x \rangle))}{\hbar} \right],
\label{eq:A5}
\end{eqnarray}
$\mu$ and $\alpha$ being time-dependent parameters.
For the introduced squeezed coherent state, equations of motion 
given by Eq. (\ref{eq:A2}) are closed 
within $\langle x \rangle$, $\langle p \rangle$,
$\langle \delta x^2\rangle$ and $\langle \delta x \delta p+\delta p \delta x\rangle$
[see Eqs. (\ref{eq:G17})-(\ref{eq:G19})].
A comparison between Heller's GWM and TDVA is made 
in Refs. \cite{Pattanayak94b,Sundaram95}.

There have been many studies on GWM which is applied to 
HO, anharmonic oscillator (AO) and Morse potentials \cite{Robinett04}. 
However, GWM has some difficulty when applied to a potential $U(x)$
including terms of $x^n$ with $n > 2$. 
Although it has been claimed that GWM yields a fairly good result 
for AO systems \cite{Cooper86}, we wonder whether it actually works 
for double-well (DW) systems. DW potential models have been employed 
in a wide range of fields including physics, chemistry and biology
(for a recent review on DW systems, see Ref. \cite{Thorwart01}).
Lin and Ballentine \cite{Lin90}, and Utermann, Dittrich and H\"{a}nggi \cite{Utermann94} 
studied semi-classical properties of DW systems 
subjected to periodic external forces, calculating the Husimi function \cite{Husimi40}.
Their calculations showed a chaotic behavior in accordance with 
classical driven DW systems.
Igarashi and Yamada \cite{Igarashi06} studied a coherent oscillation and decoherence 
induced by applied polychromatic forces in quantum DW system.
By using TDVA, Pattanayak and Schieve \cite{Pattanayak94} pointed out that
a chaos is induced by quantum noise in DW systems 
without external forces although classical counterparts are regular.
This is in contrast to the usual expectation that quantum effects suppress classical chaos. 
Chaotic-like behavior was reported in a square DW system
obtained by the exact calculation \cite{Ashkenazy95}. 
Quantum chaos pointed out in Ref. \cite{Pattanayak94}
is still controversial \cite{Pattanayak96,Blum96,Bonfim98,Bag00,Roy01,Habib04}.

Quite recently, Hasegawa has studied effects of the asymmetry 
on the specific heat \cite{Hasegawa12b} and tunneling \cite{Hasegawa13b}
in the asymmetric DW systems, 
by using the spectral method (SM) in which expansion coefficients are evaluated 
for energy matrix elements with a finite size of $N_m=30$ 
[Eqs. (\ref{eq:B7}) and (\ref{eq:B8})].
Model calculations in Refs.  \cite{Hasegawa12b,Hasegawa13b} have pointed out 
intrigue phenomena which are in contrast with earlier relevant studies.
It is worthwhile to examine the validity of TDVA applied to DW systems 
with the use of quasi-exact SM \cite{Hasegawa12b,Hasegawa13b}, 
which is the purpose of the present paper.
Such a study has not been reported as far as we are aware of.
It is important to clarify the significance of TDVA for DW systems.  
 
The paper is organized as follows.
In Section 2, we mention the calculation method employed in our study.
We consider quantum systems described by the symmetric DW (SDW) model.
In solving dynamics of a Gaussian wavepacket in the SDW, we have adopted
the two methods: SM and TDVA. In Section 3, we report calculated results
of the magnitude of wavefunction ($\vert \Psi(x,t) \vert^2$), 
an expectation value of $x$ ($\langle x \rangle$), the auto-correlation function ($C(t)$)
and the uncertainty product ($\langle \delta x^2 \rangle$$\langle \delta p^2 \rangle$). 
In Section 4 we apply our method also to an AO model. Section 5 is devoted to our conclusion.

\section{The adopted method}
\subsection{Symmetrical double-well potential}
\begin{figure}[b]
\begin{center}
\includegraphics[keepaspectratio=true,width=70mm]{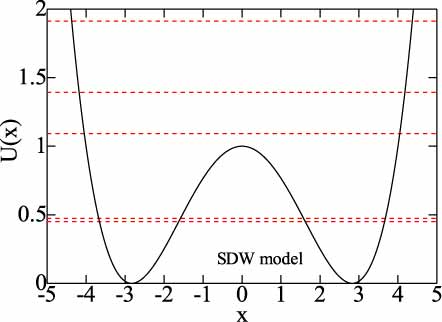}
\end{center}
\caption{
(Color online) 
The symmetric DW potential (solid curve)
with $x_s=2\sqrt{2}$ and $\Delta=1.0$ in Eq. (\ref{eq:H1}),
dashed curves expressing eigenvalues of $E_{\nu}$ ($\nu=0-4$).
}
\label{fig1}
\end{figure}

We consider a DW system whose Hamiltonian is given by \cite{Hasegawa12b,Hasegawa13b}
\begin{eqnarray}
H &=& \frac{p^2}{2 m} + U(x)=H_0+V(x),
\label{eq:B1}
\end{eqnarray}
where
\begin{eqnarray}
U(x) &=& C \;(x^2-x_s^2)^2,\;\;
\left(C=\frac{m \omega^2}{8 x_s^2} \right)
\label{eq:H1} \\
H_0 &=& \frac{p^2}{2m}+U_0(x), \\
U_0(x) &=& \frac{m \omega^2 x^2}{2}, \\
V(x) &=& U(x)-U_0(x).
\end{eqnarray}
Here $m$, $x$ and $p$ express mass, position and momentum, respectively, of a particle,
$U(x)$ stands for the DW potential, and $H_0$ is the HO Hamiltonian
with the oscillator frequency $\omega$.
The SDW potential $U(x)$
has stable minima at $x=\pm x_s$ and an unstable maximum at
$x_u=0$ with the potential barrier of $\Delta=U(0)-U(\pm x_s)=m \omega^2 x_s^2/8$.
A prefactor of $C$ in Eq. (\ref{eq:H1}) is chosen such that the DW potential $U(x)$
has the same curvature at the minima as the HO potential $U_0(x)$:
$U^{''}(\pm x_s)=U^{''}_0(0)=1.0$.
Figure \ref{fig1} expresses the adopted quartic DW potential $U(x)$ 
with $x_s=2 \sqrt{2}$ and $\Delta=1.0$ in Eq. (\ref{eq:H1}).
Eigenfunction and eigenvalue for $H_0$ are given by
\begin{eqnarray}
\phi_n(x) &=& \frac{1}{\sqrt{2^n n!}} 
\left( \frac{m \omega}{\pi \hbar} \right)^{1/4}
\exp\left( -\frac{m \omega x^2}{2 \hbar}\right)
{\cal H}_n\left( \sqrt{\frac{m \omega}{\hbar}}\:x \right), 
\label{eq:A12}\\
E_{0n} &=& \left( n+\frac{1}{2} \right) \hbar \omega
\hspace{1cm}\mbox{($n=0,1,2\cdot,\cdot\cdot\cdot$)},
\label{eq:H3}
\end{eqnarray}
where ${\cal H}_n(x)$ stands for the $n$th Hermite polynomial.

\subsection{Spectral method}
Various approximate analytical and numerical methods have been proposed to solve
the Schr\"{o}dinger equation given by Eq. (\ref{eq:A1}) \cite{Tannor07}. 
Assuming $\Psi(x,t)=\Psi(x) \:e^{-i E t/\hbar}$, we first solve the steady-state
Schr\"{o}dinger equation, $H \Psi(x)=E \Psi(x)$, with the eigenvalue $E$.
The stationary wavefunction $\Psi(x)$ is expanded in terms of $\phi_n(x)$
\begin{eqnarray}
\Psi(x) &=& \sum_{n=0}^{N_m} c_n \phi_n(x),
\label{eq:B5}
\end{eqnarray}
leading to the secular equation
\begin{eqnarray}
E c_n &=& \sum_{k=0}^{N_m} H_{nk} c_k,
\label{eq:B6}
\end{eqnarray}
with
\begin{eqnarray}
H_{nk} &=& E_{0n}\:\delta_{n,k}
+\int_{-\infty}^{\infty} \phi_n(x)^* \:V(x) \phi_k(x)\;dx,
\label{eq:B9}
\end{eqnarray}
where $N_m$ is the maximum quantum number.

For the time-dependent state, we adopt SM in which
the eigenfunction $\Psi(x,t)$ is expanded in terms of $\phi_n(x)$ with finite $N_m$
\begin{eqnarray}
\Psi(x,t) &=& \sum_{n=0}^{N_m} c_n(t) \phi_n(x).
\label{eq:B7}
\end{eqnarray}
Time-dependent expansion coefficients 
$\{ c_n(t) \}$ obey equations of motion given by
\begin{eqnarray}
i \hbar \:\frac{\partial c_n(t)}{\partial t} &=& \sum_{k=0}^{N_m} H_{nk} \:c_k(t)
\hspace{1cm}\mbox{($n=0$ to $N_m$)}.
\label{eq:B8}
\end{eqnarray}
Equation (\ref{eq:B8}) expresses the ($N_m+1$) first-order differential equations, 
which may be solved for given initial conditions of $\{ c_n(0) \}$.
Initial values of expansion coefficients $\{ c_n(0) \}$ are determined by 
\begin{eqnarray}
c_n(0) &=& \int_{-\infty}^{\infty} \phi_n(x)^* \:\Psi_G(x, 0)\:dx
\hspace{1cm}\mbox{($n=0$ to $N_m$)},
\label{eq:B10}
\end{eqnarray}
for a given Gaussian wavepacket $\Psi_G(x, 0)$ [Eq. (\ref{eq:A5})]
\begin{eqnarray}
\Psi_G(x, 0)=\frac{1}{(2 \pi \mu_0)^{1/4}} \:\exp\left[-\frac{(1-i \alpha_0)}{4 \mu_0}(x-x_0)^2
+ i \:\frac{p_0 (x-x_0)}{\hbar} \right],
\label{eq:B11}
\end{eqnarray}
where $x_0$ and $p_0$ are initial position and momentum, respectively, and
$\mu_0$ and $\alpha_0$ are assumed initial parameters at $t=0.0$.
Once solutions of $\{ c_n(t) \}$ in Eq. (\ref{eq:B8}) are obtained, 
the wavefunction $\Psi(x,t)$ may be constructed by Eq. (\ref{eq:B7}).
 
Matrix elements $H_{n k}$ in Eq. (\ref{eq:B9}) may be analytically evaluated, and
various time-dependent averages such as $\langle x \rangle$ and  $\langle p \rangle$
are expressed in terms of $\{ c_n(t) \}$ (see the Appendix). We expect that SM 
with $N_m=30$ adopted in our numerical calculations is fairly accurate \cite{Hasegawa12b,Hasegawa13b}. 
Some results of SM have been cross-checked, by solving the Schr\"{o}dinger equation 
with the use the MATHEMATICA resolver for the partial differential equation.  

\subsection{Time-dependent variational approximation}
Equations of motion in Eq. (\ref{eq:A2}) 
are expressed by 
\begin{eqnarray}
\frac{d \langle x \rangle}{dt} &=& \frac{\langle p \rangle}{m}, 
\label{eq:G1}
\\
\frac{d \langle p \rangle}{d t} &=& - U'(\langle x \rangle) 
- \sum_{k=2}^{\infty} \frac{U^{(k+1)}(\langle x \rangle)}{k !} \langle \delta x^k \rangle, 
\label{eq:G2}
\\
\frac{d \langle \delta x^2 \rangle}{dt} 
&=& \frac{1}{m} \langle \delta x \delta p
+\delta p \delta x \rangle, 
\\
\frac{d \langle  \delta x \delta p +\delta p \delta x  \rangle}{dt}
&=& -2 \sum_{k=1}^{\infty} \frac{U^{(k+1)}(\langle x \rangle)}{k !} \langle \delta x^{k+1} \rangle
+ \frac{2}{m} \langle \delta p^2 \rangle, 
\label{eq:G3}\\
\frac{d \langle \delta p^2 \rangle}{dt} 
&=& 
- \sum_{k=1}^{\infty} \frac{U^{(k+1)}(\langle x \rangle)}{k !}
\langle \delta x^{k} \delta p+\delta p \delta x^{k} \rangle.
\label{eq:G4} 
\end{eqnarray}
Equations (\ref{eq:G1})-(\ref{eq:G4}) include higher-order fluctuations 
which are not closed in general.
It is possible to construct various approximations depending on
how many terms are taken into account in Eqs. (\ref{eq:G1})-(\ref{eq:G4}).
If we neglect the second term of Eq. (\ref{eq:G2}),
Eqs. (\ref{eq:G1}) and (\ref{eq:G2}) form classical equations of motion.
When we neglect the second term in Eq. (\ref{eq:G2}) 
and truncate Eqs. (\ref{eq:G3}) and (\ref{eq:G4})
at $k=1$, Eqs. (\ref{eq:G1})-(\ref{eq:G4}) reduce to equations of motion in Heller's GWM.
Equations of motion including up to fourth-order corrections were obtained
in Ref. \cite{Sundaram95}.

To close a hierarchal chain of equations of motion, TDVA assumes that a wavepacket 
is expressed by the normalized squeezed coherent state given by Eq. (\ref{eq:A5}),
implying relations \cite{Cooper86,Pattanayak94,Pattanayak94b,Sundaram95}
\begin{eqnarray}
\langle \delta x^{2 \ell} \rangle 
&=& \frac{(2 \ell)!}{\ell ! \:2^{\ell}} \:\mu^{\ell}, 
\;\;\;\;
\langle \delta x^{2 \ell+1} \rangle = 0,
\hspace{0.5cm} \mbox{$(\ell= 1,2,\cdot\cdot)$}
\label{eq:G12} \\
\langle \delta p^2 \rangle &=& \frac{\hbar^2+\alpha^2}{4 \mu},
\label{eq:G13a} \\
\langle \delta x \delta p +\delta p \delta x \rangle &=& \alpha, 
\label{eq:G13b}
\end{eqnarray}
where $\mu$ and $\alpha$ are time-dependent parameters.
Note that Eqs. (\ref{eq:G12})-(\ref{eq:G13b}) yield the uncertainty product expressed by
\begin{eqnarray}
\langle \delta x^2 \rangle \langle \delta p^2 \rangle 
&=& \frac{\hbar^2+ \langle \delta x \delta p +\delta p \delta x \rangle^2}{4}.
\label{eq:G13}
\end{eqnarray}
These lead to equations of motion given by
\begin{eqnarray}
\frac{d \langle x \rangle}{dt} &=& \frac{\langle p \rangle }{m}, 
\label{eq:G5a}\\
\frac{d \langle p \rangle}{d t} 
&=& 
- U'(\langle x \rangle)
- \sum_{\ell=1}^{\infty} \frac{U^{(2 \ell+1)}(\langle x \rangle)}{\ell !\;2^{\ell}} \:\mu^{\ell}, 
\label{eq:G5b}\\
\frac{d \mu}{dt} &=& \frac{\alpha}{m}, 
\label{eq:G5d}\\
\frac{d \alpha}{dt} &=& \frac{\hbar^2+\alpha^2}{2 m \mu}
- \sum_{\ell=1}^{\infty} \frac{U^{(2 \ell)}(\langle x \rangle)}{(\ell-1)! 
\;2^{\ell-2}} \:\mu^{\ell}.
\label{eq:G5c}
\end{eqnarray}
Alternatively, Eqs. (\ref{eq:G5a})-(\ref{eq:G5c}) may be rewritten as
%
\begin{eqnarray}
\frac{d \langle x \rangle}{dt} &=& \frac{\langle p \rangle }{m}, 
\label{eq:G17}\\
\frac{d \langle p \rangle}{d t} 
&=& 
- U'(\langle x \rangle)
- \sum_{\ell=1}^{\infty} \frac{U^{(2 \ell+1)}(\langle x \rangle)  
\langle \delta x^2 \rangle^{\ell}}{\ell !\;2^{\ell}}, 
\label{eq:G18}\\
\frac{d \langle \delta x^2 \rangle}{dt} 
&=& \frac{1}{m}\langle \delta x \delta p +\delta p \delta x \rangle, \\
\frac{d \langle \delta x \delta p +\delta p \delta x \rangle}{dt} 
&=& \frac{\hbar^2+\langle \delta x \delta p +\delta p \delta x \rangle^2}{2 m \langle \delta x^2 \rangle}
- \sum_{\ell=1}^{\infty} \frac{U^{(2 \ell)}
(\langle x \rangle) \langle \delta x^2 \rangle^{\ell}}{(\ell-1)! \: 2^{\ell-2}},
\label{eq:G19}
\end{eqnarray}
which show a closure of equations of motion within $\langle x \rangle$, $\langle p \rangle$,
$\langle \delta x^2 \rangle$ and $\langle \delta x \delta p +\delta p \delta x \rangle$.

\section{Model calculations}

\begin{figure}
\begin{center}
\includegraphics[keepaspectratio=true,width=70mm]{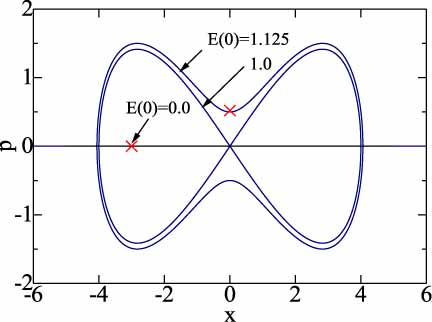}
\end{center}
\caption{
(Color online) 
The classical $x$-$p$ phase space of the SDW model
for various initial energies of $E(0)$, marks $\times $ denoting positions 
of initial states adopted in model calculations.
}
\label{fig2}
\end{figure}

We apply our calculation method to the SDW potential given by Eq. (\ref{eq:H1}).
We have calculated energy matrix elements of $H_{kn}$, by using
Eq. (\ref{eq:H6}) with $m=\omega=\hbar=1.0$ and $N_m=30$.
Obtained eigenvalues  are $E_{\nu}=$ 
0.450203, 0.474126, 1.09262, 1.39334 and 1.91286
for $\nu=0$ to 4, respectively, which are plotted by dashed curves in Fig. \ref{fig1}.
The ground state ($E_0$) and first excited state ($E_1$), which are quasi-degenerate, 
are below the potential barrier of $\Delta=1.0$.
The energy gap between ground and first excited states is 
$\Delta E = E_1-E_0 = 0.023923$. 
Low-lying eigenvalues calculated with $N_m=30$ are in good agreement
with those obtained with $N_m=20$ \cite{Hasegawa12b}.

Figure \ref{fig2} shows the classical $x$-$p$ phase space for initial energies
of $E(0)=0.0$, 1.0 and 1.125.
Marks $\times$ in Fig. \ref{fig2} show two initial states in the $x$-$p$ phase space
adopted in our calculations. Calculated results for the two initial states of 
$(x_0, p_0)=(-2\sqrt{2}, 0.0)$ and $(x_0, p_0)=(0.0, 0.5)$
will be separately reported in the following.

\subsubsection{Case of the initial state of $(x_0,p_0)=(-2\sqrt{2}, 0.0)$}

We have adopted the Gaussian wavepacket $\Psi_G(x, 0)$ 
locating at the stable point of the left well with
$(x_0,p_0)=(-2\sqrt{2}, 0.0)$, and $\mu_0=0.1$ and $\alpha_0=0.0$ at $t=0.0$,
which yields the minimum uncertainty product of
$\langle \delta x^2 \rangle \langle \delta p^2\rangle=1/4$. 
Initial coefficients $\{ c_n(0) \}$ calculated by Eq. (\ref{eq:B10}) are real with
appreciable magnitudes for $3 \lesssim n \lesssim 10$.
A norm of the initial Gaussian wavepacket is 
$\sum_n c_n(0)^* c_n(0)=0.999999$ [Eq. ({\ref{eq:C2}})].
After solving ($N_m+1$) first-order differential equations for $\{ c_n(t) \}$ 
given by Eq. (\ref{eq:B8}) for initial values of $\{ c_n(0) \}$, 
we obtain the time-dependent eigenfunction $\Psi(x,t)$ expressed in terms of $\{ c_n(t) \}$ in Eq. (\ref{eq:B7}). 

\begin{figure}
\begin{center}
\includegraphics[keepaspectratio=true,width=120mm]{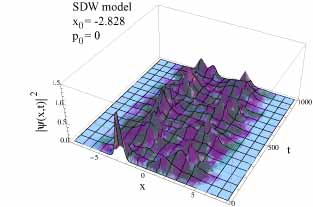}
\end{center}
\caption{
(Color online) 
3D plot of $\vert \Psi(x,t) \vert^2$ as functions of $x$ and $t$
calculated by SM for the SDW model ($x_0=-2\sqrt{2}, p_0=0.0$). 
}
\label{fig3}
\end{figure}

\begin{figure}
\begin{center}
\includegraphics[keepaspectratio=true,width=100mm]{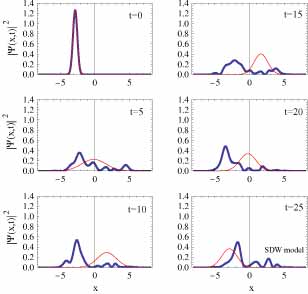}
\end{center}
\caption{
(Color online) 
The $x$ dependence of $\vert \Psi(x,t) \vert^2$ at various $t$ of the SDW model
calculated by SM (bold solid curves) and TDVA (solid curve) ($x_0=-2\sqrt{2}, p_0=0.0$).
}
\label{fig4}
\end{figure}

Figure \ref{fig3} shows the 3D plot of $\vert \Psi(x,t) \vert^2$
calculated by SM. 
We note that the Gaussian wavepacket in SM quickly spreads as the time develops.
In order to scrutinize the behavior of $\vert \Psi(x,t) \vert^2$ at small $t$,
its time dependence at $0 \leq t \leq 25$ is plotted by bold solid curves in Fig. \ref{fig4},
where solid curves denote results of TDVA.
The Gaussian wavepacket becomes widespread even at $t=5.0$ in SM, and its 
trend becomes more significant with increasing $t$.
Figure \ref{fig4} clearly shows that $\vert \Psi(x,t) \vert^2$ in SM is quite different 
from that in TDVA and that $\Psi(x,t)$ cannot be expressed 
by a single Gaussian except at $t=0.0$.

\begin{figure}
\begin{center}
\includegraphics[keepaspectratio=true,width=70mm]{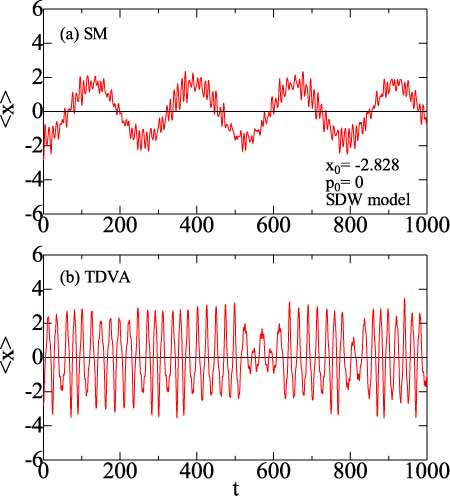}
\end{center}
\caption{
(Color online) 
The time dependence of $\langle x \rangle$ of the SDW model calculated 
by (a) SM and (b) TDVA ($x_0=-2\sqrt{2}, p_0=0.0$).
}
\label{fig5}
\end{figure}

\begin{figure}[htbp]
\begin{center}
\includegraphics[keepaspectratio=true,width=70mm]{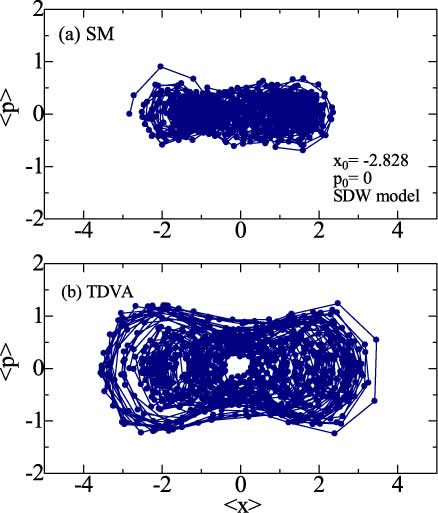}
\end{center}
\caption{
(Color online) 
The $\langle x \rangle$ vs. $\langle p \rangle$ plot of the SDW model 
calculated by (a) SM and (b) TDVA, time step being $\Delta t=1.0$ for $0 \leq t < 1000$
($x_0=-2\sqrt{2}, p_0=0.0$).
}
\label{fig6}
\end{figure}

\begin{figure}
\begin{center}
\includegraphics[keepaspectratio=true,width=70mm]{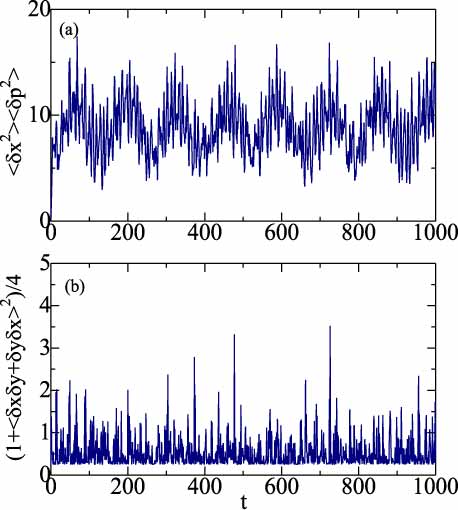}
\end{center}
\caption{
(Color online) 
Time dependences of (a) the uncertainty product of
$\langle \delta  x^2 \rangle \langle \delta  p^2 \rangle$
and (b) $(1+\langle \delta x \delta p+ \delta x \delta p \rangle^2)/4$ 
of the SDW model calculated by SM ($x_0=-2\sqrt{2}, p_0=0.0$).
Note that  $\langle \delta  x^2 \rangle \langle \delta  p^2 \rangle$
equals to $(1+\langle \delta x \delta p+ \delta x \delta p \rangle^2)/4$
in TDVA [Eq. (\ref{eq:G13})], 
which is not realized in (a) and (b).
}
\label{fig7}
\end{figure}

\begin{figure}
\begin{center}
\includegraphics[keepaspectratio=true,width=70mm]{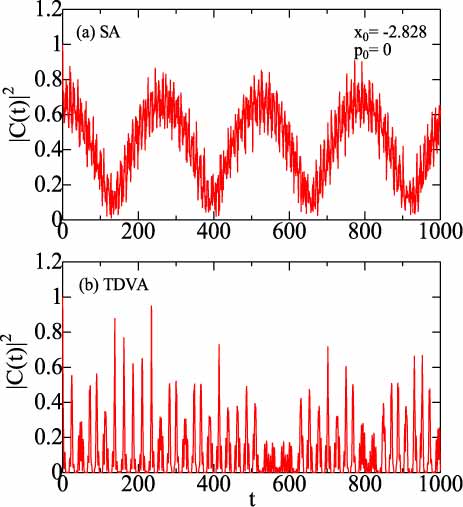}
\end{center}
\caption{
(Color online) 
Time dependences of the auto-correlation function
$\vert C(t) \vert^2$ of the SDW model calculated by (a) SM and (b) TDVA
($x_0=-2\sqrt{2},p_0=0.0$).
}
\label{fig8}
\end{figure}

The difference between SM and TDVA is more clearly seen in the
time-dependent expectation value of $\langle x \rangle$.
Figure \ref{fig5}(a) shows $\langle x \rangle$ of SM expressing a tunneling of a particle 
with the period of about 260, which is consistent with the period
estimated from the energy gap by $T = 2 \pi/\Delta E=262$. 
On the contrary, $\langle x \rangle$ of TDVA in Fig. \ref{fig5}(b)
shows more rapid oscillation with a period of about $25 - 30$.

Figures \ref{fig6}(a) and \ref{fig6}(b) show $\langle x \rangle$ vs.
$\langle p \rangle$ plots calculated by SM and TDVA, respectively.
The $\langle x \rangle$ vs. $\langle p \rangle$ plot of SM in Fig. \ref{fig6}(a)
is quite different from that of TDVA in Fig. \ref{fig6}(b). 

Figure \ref{fig7}(a) shows the uncertain product of
$\langle \delta x^2\rangle \langle \delta p^2\rangle$ calculated by SM,
which expresses a measure of quantum fluctuation.
It starts from the minimum uncertainty of $1/4$ at $t=0$, 
and with increasing $t$ it grows and oscillates 
between about 5 and 17 with the period of about 130. For a comparison, we plot
$(1+\langle \delta x \delta p+ \delta x \delta p \rangle^2)/4$
in Fig. \ref{fig7}(b).  
TDVA assumes the equality of $\langle \delta x^2\rangle \langle \delta p^2\rangle
=(1+\langle \delta x \delta p+ \delta x \delta p \rangle^2)/4$ as given by Eq. (\ref{eq:G13}). 
Figures \ref{fig7}(a) and \ref{fig7}(b), however, imply that this equality is not satisfied in SM.

Figures \ref{fig8}(a) and \ref{fig8}(b) show the auto-correlation functions 
$\vert C(t) \vert^2$ calculated by SM and TDVA, respectively, with Eq. (\ref{eq:C2})
in the Appendix.
$\vert C(t) \vert^2$ of SM, which is unity at $t=0.0$, oscillates between about
0.1 and 0.7 with a period of about 260.
The result of SM in Fig. \ref{fig8}(a) is again quite different from that of TDVA
in Fig. \ref{fig8}(b).

\subsubsection{Case of the initial state of $(x_0,p_0)=(0.0,0.5)$}
\begin{figure}
\begin{center}
\includegraphics[keepaspectratio=true,width=110mm]{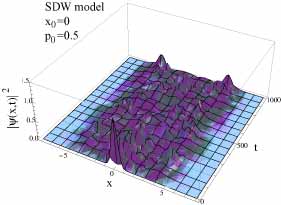}
\end{center}
\caption{
(Color online) 
3D plot of $\vert \Psi(x,t) \vert^2$ as functions of $x$ and $t$
calculated by SM for the SDW model ($x_0=0.0, p_0=0.5$). 
}
\label{fig9}
\end{figure}

\begin{figure}
\begin{center}
\includegraphics[keepaspectratio=true,width=100mm]{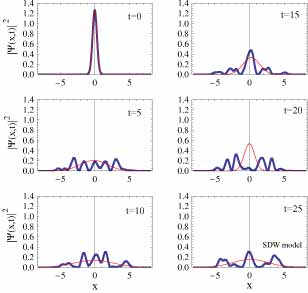}
\end{center}
\caption{
(Color online) 
The $x$ dependence of $\vert \Psi(x,t) \vert^2$ at various $t$ of the SDW model
calculated by SM (bold solid curves) and TDVA (solid curves) ($x_0=0.0, p_0=0.5$).
}
\label{fig10}
\end{figure}

\begin{figure}
\begin{center}
\includegraphics[keepaspectratio=true,width=70mm]{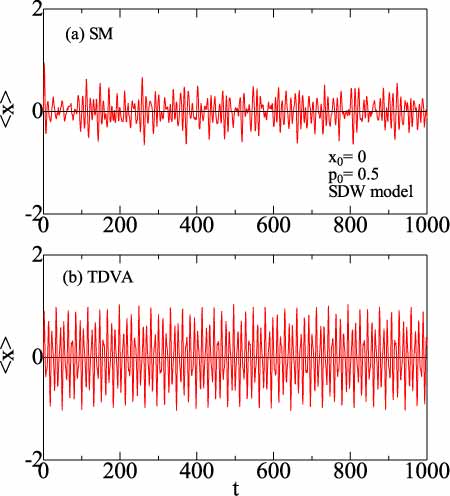}
\end{center}
\caption{
(Color online) 
Time dependences of $\langle x \rangle$ of the SDW model
calculated by (a) SM and (b) TDVA ($x_0=0.0, p_0=0.5$).
}
\label{fig11}
\end{figure}

Next we adopt a Gaussian wavepacket with a different initial state of
$(x_0, p_0)=(0.0, 0.5)$ but with the same $\mu_0=0.1$ and $\alpha_0=0.0$ at $t=0.0$. 
Initial coefficients $\{ c_n(0) \}$ calculated by Eq. (\ref{eq:B10}) are complex with 
appreciable magnitudes for $0 \lesssim n \lesssim 15$.
The initial state of $(x_0,p_0)=(0.0,0.5)$ locates near a top of
the potential barrier (see Fig. \ref{fig2}).
Note that in the classical calculation, the $x$ vs. $p$ plot
forms a cocoon shape extending from $x=-4.06021$ to $x=4.06021$
and from $p=-1.5$ to 1.5, as shown in Fig. \ref{fig2}.
Then at $t > 0$, a particle starting from $(x_0, p_0)=(0.0, 0.5)$
rolls down the potential up to $x=4.06021$
and then approaches $x=-4.06021$ after passing through $x=0$ in the classical calculation.
However, this classical behavior is quite different from
quantum results calculated by SM and TDVA.  
The 3D plot of $\vert \Psi(x,t) \vert^2$ of SM shown in Fig. \ref{fig9} 
has appreciable magnitudes at $-5 \lesssim x \lesssim 5$ for $0 < t < 1000$.
Bold solid curves and solid curves in Fig. \ref{fig10} show
$\vert \Psi(x,t) \vert^2$ calculated by SM and TDVA, respectively.
$\vert \Psi(x,t) \vert^2$ of SM, which is distorted and spreads at $t > 0$, 
is different from the relevant result of TDVA.
An expectation value of $\langle x \rangle$ of SM in Fig. \ref{fig11}(a)
does not so much depart from the initial point of $x=0.0$ 
in contrast to that of TDVA shown in Fig. \ref{fig11}(b).

\begin{figure}
\begin{center}
\includegraphics[keepaspectratio=true,width=70mm]{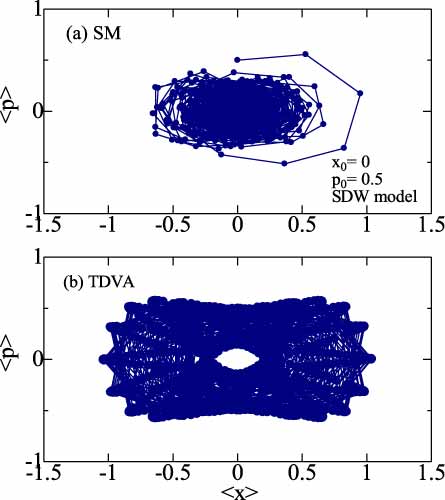}
\end{center}
\caption{
(Color online) 
The $\langle x \rangle$ vs. $\langle p \rangle$ plot of the SDW model
calculated by (a) SM and (b) TDVA,
time step being $\Delta t=1.0$ for $0 \leq t < 1000$ ($x_0=0.0, p_0=0.5$).
}
\label{fig12}
\end{figure}

Figures \ref{fig12}(a) and \ref{fig12}(b) show 
$\langle x \rangle$ vs. $\langle p \rangle$ plots calculated
by SM and TDVA, respectively. 
The result of SM in Fig. \ref{fig12}(a) exhibits a random-like motion, 
which is different from a quasi-periodic motion  of TDVA in Fig. \ref{fig12}(b).

\begin{figure}
\begin{center}
\includegraphics[keepaspectratio=true,width=70mm]{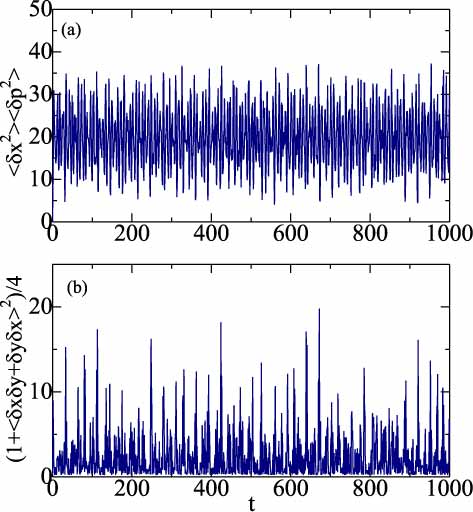}
\end{center}
\caption{
(Color online) 
Time dependences of (a) the uncertainty product of
$\langle \delta  x^2 \rangle \langle \delta  p^2 \rangle$
and (b) $(1+\langle \delta x \delta p+ \delta x \delta p \rangle^2)/4$ 
of the SDW model calculated by SM ($x_0=0.0, p_0=0.5$).
Note that  $\langle \delta x^2 \rangle \langle \delta  p^2 \rangle$
equals to $(1+\langle \delta x \delta p+ \delta x \delta p \rangle^2)/4$
in TDVA [Eq. (\ref{eq:G13})], which is not held in SM.
}
\label{fig13}
\end{figure}

Figures \ref{fig13}(a) and \ref{fig13}(b) show
$\langle \delta  x^2 \rangle \langle \delta  p^2 \rangle$ and
$(1+\langle \delta x \delta p+ \delta x \delta p \rangle^2)/4$, respectively,
calculated by SM.
We note that
$\langle \delta  x^2 \rangle \langle \delta  p^2 \rangle
\neq (1+\langle \delta x \delta p+ \delta x \delta p \rangle^2)/4$ 
in SM, which is in contrast with Eq. (\ref{eq:G13}) in TDVA.

\section{Discussion}
\subsection{An effective Hamiltonian in TDVA}
Refs. \cite{Cooper86,Pattanayak94,Pattanayak94b} showed that
by using a change of variables given by 
\begin{eqnarray}
\rho^2 &=& \langle \delta x^2 \rangle, 
\label{eq:G20}\\
\pi &=& \frac{1}{2 \rho}
\langle \delta x \delta p +\delta p \delta x \rangle,
\label{eq:G21}
\end{eqnarray}
Eqs. (\ref{eq:G5a})-(\ref{eq:G5c}) are transformed to
\begin{eqnarray}
\frac{d \langle x \rangle}{dt} &=& \frac{\langle p \rangle }{m}, 
\label{eq:G22}\\
\frac{d \langle p \rangle}{d t} 
&=& 
- U'(\langle x \rangle)- \sum_{\ell=1}^{\infty} \frac{U^{(2 \ell+1)}(\langle x \rangle)}{\ell !\;2^{\ell}} \rho^{2 \ell}, 
\label{eq:G23}\\
\frac{d \rho}{dt} &=& \frac{\pi}{m}, \\
\frac{d \pi}{dt} &=& \frac{\hbar^2}{4 m \rho^3}
- \sum_{\ell=1}^{\infty} \frac{U^{(2 \ell)}(\langle x \rangle)}{(\ell-1)! \;2^{\ell-1}} \rho^{2 \ell-1}.
\label{eq:G24}
\end{eqnarray}
It was shown that fluctuation variables $\rho$ and $\pi$ are 
conjugate and that the effective Hamiltonian may be expressed in the extended phase space
spanned by $\langle x \rangle$, $\langle p \rangle$, $\rho$ and $\pi$ as given by
\cite{Cooper86,Pattanayak94,Pattanayak94b}
\begin{eqnarray}
H_{eff} &=& \frac{\langle p\rangle^2}{2m}+\frac{\pi^2}{2m}+\frac{\hbar^2}{8 m \rho^2} 
+ U(\langle x \rangle)
+ \sum_{\ell=1}^{\infty} \frac{U^{(2 \ell)}(\langle x \rangle)}{\ell ! \:2^{\ell}} \rho^{2 \ell}. 
\label{eq:G25}
\end{eqnarray}
We should note that the effective Hamiltonian given by Eq. (\ref{eq:G25})
relies on the identities given by Eqs. (\ref{eq:G12})-(\ref{eq:G13b})
which are based on the assumed squeezed Gaussian wavepacket 
given by Eq. (\ref{eq:A5}).
If these identities are not held as our SM calculation suggests, 
the effective Hamiltonian given by Eq. (\ref{eq:G25}) is not valid in DW systems. 

\subsection{Anharmonic Oscillator}

\begin{figure}
\begin{center}
\includegraphics[keepaspectratio=true,width=100mm]{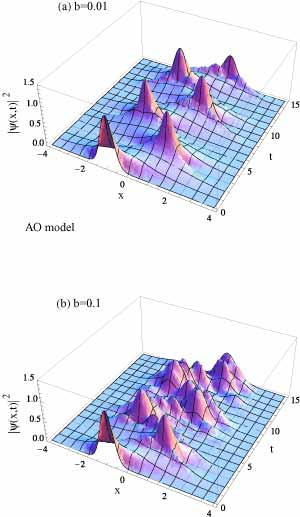}
\end{center}
\caption{
(Color online) 
3D plot of $\vert \Psi(x,t) \vert^2$ as functions of $x$ and $t$ of the AO model
with (a) $b=0.01$ and (b) $b=0.1$ calculated by SM ($x_0=-1.0,p_0=0.0$).
}
\label{fig14}
\end{figure}

\begin{figure}
\begin{center}
\includegraphics[keepaspectratio=true,width=100mm]{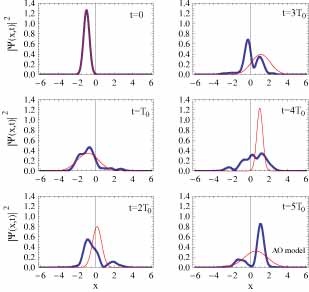}
\end{center}
\caption{
(Color online) 
The $x$ dependence of $\vert \Psi(x,t) \vert^2$ at $t=k T_0$ ($k=0,1,2 \cdots$)
of the AO model with $b=0.1$ calculated by SM (bold solid curves) and TDVA (solid curve)
($x_0=-1.0,p_0=0.0$), where $T_0=6.283$ $(=2 \pi/\omega)$ is a period for a HO system ($b=0$).
}
\label{fig15}
\end{figure}

\begin{figure}
\begin{center}
\includegraphics[keepaspectratio=true,width=70mm]{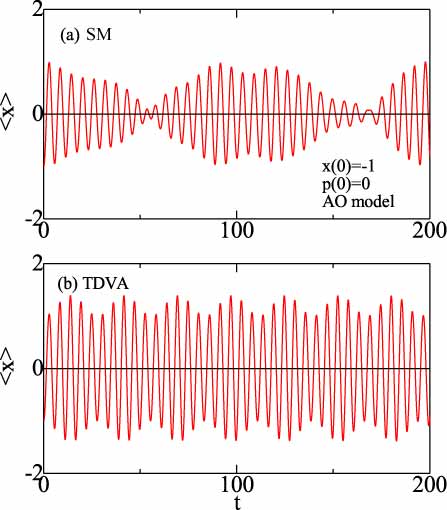}
\end{center}
\caption{
(Color online) 
Time dependences of $\langle x \rangle$ in the AO model with $b=0.1$
calculated by (a) SM and (b) TDVA ($x_0=-1.0,p_0=0.0$).
}
\label{fig16}
\end{figure}

We have studied Gaussian wavepacket dynamics of quantum DW systems
in the preceding section.
It is worthwhile to examine also an AO model 
given by
\begin{eqnarray}
U(x) &=& \frac{x^2}{2}+ \frac{b x^4}{4}=U_0(x)+ \frac{b x^4}{4},
\label{eq:J1}
\end{eqnarray}
where $b$ expresses a degree of anharmonicity.
We have repeated calculations, by using SM and TDVA with necessary modifications.

Figure \ref{fig14}(a) shows the 3D plot of $\vert \Psi(x,t) \vert^2$ for $b=0.01$
with an assumed Gaussian state for $(x_0,p_0)=(-1.0,0.0)$, and $\mu_0=0.1$
and $\alpha_0=0.0$ at $t=0.0$. 
In the case of a HO potential ($b=0.0$), $\vert \Psi(x,t) \vert^2$
is periodic with a period of $T_0=2 \pi/\omega= 6.283 $. 
It is shown in \ref{fig14}(a) that $\vert \Psi(x,t) \vert^2$ 
for a small $b=0.01$ is nearly periodic as that for $b=0.0$ at $t < 10 $.
For a larger $b=0.1$, however, this periodicity is destroyed
and the wavepacket spreads in the non-Gaussian form as Fig. \ref{fig14}(b) shows.
This is more clearly realized in Fig. \ref{fig15} where
$\vert \Psi(x,t) \vert^2$ calculated by SM (bold solid curves) 
are quite different from their counterparts obtained by TDVA (solid curve).
Our result of SM in Fig. \ref{fig15} is consistent with that in Ref. \cite{Herrera12}
which studied effects of anharmonicity and interactions in DW systems.

Figures \ref{fig16}(a) and \ref{fig16}(b) show time dependences of  
$\langle x \rangle$ calculated by SM and TDVA, respectively.
$\langle x \rangle$ oscillates with a period of about $5.6$ in both results.
However, a period of its envelope variation in SM ($\sim 110$) is
larger than that in TDVA ($\sim 30$): the former corresponds to
the revival time after which a wavepacket periodically returns to the initial shape.

\section{Conclusion}
By using SM and TDVA, we have calculated time dependences of wavefunctions, 
averages of position and momentum, auto-correlation function, 
and uncertainty product in quantum SDW systems. 
The validity of TDVA has been examined by comparisons between results of SM and TDVA.
We have obtained following results:

\vspace{0.2cm}
\noindent
(1) An initial Gaussian wavepacket in DW systems of SM spreads and deforms at $t > 0$ where
a wavepacket cannot be expressed by a single Gaussian in contrast to TDVA,

\noindent
(2) Time dependences of expectation values of $\langle x \rangle$ and 
$\langle p \rangle$, and the auto-correlation function 
in SM are quite different from their counterparts in TDVA, and

\noindent
(3) The identity relation for uncertainty product assumed in TDVA [Eq. (\ref{eq:G13})]
is not satisfied in SM.

\noindent
The item (1) holds also in asymmetric DW systems \cite{Hasegawa13b}.
The item (2) implies that the tunneling phenomenon characteristic in DW systems 
cannot be well accounted for in TDVA (Fig. \ref{fig5}) \cite{Hasegawa13b}.
The item (3) suggests that the effective Hamiltonian in the extended phase space given 
by Eq. (\ref{eq:G25}) does not hold in DW systems because it is derived with
the squeezed Gaussian wavepacket with assumptions given 
by Eqs. (\ref{eq:G12})-(\ref{eq:G13b}) in TDVA. 
GWM is best applied to dynamics in HO and AO with a small anharmonicity,
for which it provides us with an efficient and
physically-transparent calculation method.
Our calculations, however, point out that GWM is not a good approximation for DW systems. 
For a better description of quantum DW systems, it might be necessary to
adopt extended GWMs with superimposed multiple Gaussian wavepackets
(see Ref. \cite{Zoppe05}, related references therein),
which are much sophisticated and complicated than the original Heller's GWM \cite{Heller75}.

\begin{acknowledgments}
This work is partly supported by
a Grant-in-Aid for Scientific Research from 
Ministry of Education, Culture, Sports, Science and Technology of Japan.  
\end{acknowledgments}

\appendix*

\section{Matrix elements and various expectation values}
\renewcommand{\theequation}{A\arabic{equation}}
\setcounter{equation}{0}

Matrix elements $H_{nk}$ in Eq. (\ref{eq:B9}) are given as follows:
We rewrite the potential $U(x)$ as
\begin{eqnarray}
U(x) &=& \frac{A_4 x^4}{4}+\frac{A_3 x^3}{3}+\frac{A_2 x^2}{2}+A_1 x+A_0,
\label{eq:H4}
\end{eqnarray}
with
\begin{eqnarray}
A_4 &=&\frac{m \omega^2}{2 x_s^2}, \;\;A_3=0, \;\;A_2=-\frac{m\omega^2}{2},\;\;
A_1=0,\;\;A_0=\frac{m \omega^2 x_s^2}{8}.
\label{eq:H5}
\end{eqnarray}
By using relations given by 
\begin{eqnarray}
q &=& \sqrt{\frac{g}{2}}(a^{\dagger}+a),\;\;\;
p = i \frac{\hbar}{\sqrt{2 g}} (a^{\dagger}- a),
\hspace{1cm} \mbox{$\left( g=\frac{\hbar}{m \omega} \right)$}
\label{eq:C1a}\\
a^{\dagger} \:\phi_n &=& \sqrt{n+1} \:\phi_{n+1},\;\; 
a \:\phi_n = \sqrt{n} \:\phi_{n-1}, 
\label{eq:C1b}
\end{eqnarray}
we obtain the symmetric matrix elements $H_{nk}$ for $n \geq k$ given by 
\begin{eqnarray}
H_{nk} &=& \left[ \left( n+1/2\right) \hbar \omega 
+ \frac{3 A_4 g^2}{16}(2 n^2+2 n +1) +\frac{A_2' \:g}{2}(n+1/2)
+ A_0 \right] \:\delta_{n,k} \nonumber \\
&+& \left[A_3 \left(\frac{g}{2}\right)^{3/2} n \sqrt{n} 
+ A_1 \left( \frac{g}{2} \right)^{1/2} \sqrt{n} \right] \delta_{n-1,k} \nonumber \\
&+&  \left[ \frac{A_4 g^2}{8} (n-1)\sqrt{n(n-1)} 
+ \frac{A_2' \:g}{4} \sqrt{n(n-1)} \right] \:\delta_{n-2,k}  \nonumber \\
&+& \frac{A_3}{3} \left(\frac{g}{2}\right)^{3/2} \sqrt{n(n-1)(n-2)}\: \delta_{n-3,k}
\nonumber \\
&+&  \frac{A_4 g^2}{16} \sqrt{n(n-1)(n-2)(n-3)} \:\delta_{n-4,k},
\label{eq:H6}
\end{eqnarray}
where $A_2'=A_2- m \omega^2$.

Various time-dependent quantities may be expressed in terms of
$\{ c_n(t) \}$ as follows:
After some manipulations with the use of the relations given by
Eqs.(\ref{eq:C1a}) and (\ref{eq:C1b}),
the auto-correlation function is given by
\begin{eqnarray}
C(t) &=& \int_{-\infty}^{\infty} \Psi(x,t)^* \Psi(x,0) \:dx, \\
&=& \sum_{n=0}^{N_m} c_n(t)^*\:c_n(0),
\label{eq:C2}
\end{eqnarray}
and expectation values such as $x(t)$ and $p(t)$ are expressed by
\begin{eqnarray}
\langle x(t) \rangle 
&=& \sqrt{\frac{g}{2}}
\sum_n \left[\sqrt{n+1}\: c_{n+1}^*(t) c_n(t)+ \sqrt{n}\: c_{n-1}^*(t) c_n(t) \right], 
\label{eq:C3a} \\
\langle p(t) \rangle 
&=& i \sqrt{\frac{\hbar^2}{2 g}}
\sum_n \left[\sqrt{n+1}\: c_{n+1}^*(t) c_n(t) - \sqrt{n}\: c_{n-1}^*(t) c_n(t) \right], \\
\langle x(t)^2 \rangle
&=& \left( \frac{g}{2} \right)
\sum_n [ \sqrt{(n+1)(n+2)} \:c_{n+2}^*(t) c_n(t)+(2 n+1) \:c_{n}^*(t) c_n(t) 
\nonumber \\
&+& \sqrt{n(n-1)} \:c_{n-2}^*(t) c_n(t)], \\
\langle p(t)^2 \rangle &=& - \left( \frac{\hbar^2}{2 g} \right)
\sum_n [\sqrt{(n+1)(n+2)} \:c_{n+2}^*(t) c_n(t)-(2 n+1) \:c_{n}^*(t) c_n(t) 
\nonumber \\
&+& \sqrt{n(n-1)} \:c_{n-2}^*(t) c_n(t)], \\
\langle x(t)p(x)+p(t)x(t) \rangle
&=& i \:\hbar \sum_n [ \sqrt{(n+1)(n+2)} \:c_{n+2}^*(t) c_n(t) \nonumber \\
&-& \sqrt{n(n-1)} \:c_{n-2}^*(t) c_n(t)]. 
\label{eq:C3b}
\end{eqnarray}


\end{document}